\documentclass[cameraready]{Interspeech}
\title{Speech Recognition on TV Series with Video-Guided Post-ASR Correction}

\author[affiliation={1,2}, equalcontribution]{Haoyuan}{Yang}
\author[affiliation={2}, equalcontribution]{Yue}{Zhang}
\author[affiliation={2}]{Liqiang}{Jing}
\author[affiliation={1}, correspondingauthor]{John}{Hansen}
\address{
    $^1$ Center for Robust Speech Systems (CRSS), The University of Texas at Dallas, USA \\
    $^2$ Department of Computer Science, The University of Texas at Dallas, USA 
}

\email{haoyuan.yang@utdallas.edu, yue.zhang@utdallas.edu, jingliqiang6@gmail.com, john.hansen@utdallas.edu}


\usepackage{comment}

\usepackage{amsmath,graphicx,hyperref}
\usepackage{booktabs}   % \toprule, \midrule, \cmidrule, \bottomrule
\usepackage{multirow}   % \multirow
\usepackage{graphicx}   % \resizebox
\usepackage{pifont}     % \ding for ✓ and ✗

\usepackage{pgfplots}
\pgfplotsset{compat=1.17}
\usepackage{colortbl,xcolor}
\usepackage[table]{xcolor}  % in preamble
\usepackage{makecell}

\usepackage{array}  % in preamble
\definecolor{lightred}{RGB}{255,200,200}
\definecolor{lightgreen}{RGB}{200,255,200}

\definecolor{g1}{RGB}{240,249,245}
\definecolor{g2}{RGB}{224,243,234}
\definecolor{g3}{RGB}{209,237,223}
\definecolor{g4}{RGB}{193,231,212}
\definecolor{g5}{RGB}{177,225,201}
\definecolor{g6}{RGB}{162,219,190}
\definecolor{g7}{RGB}{146,213,179}
\definecolor{g8}{RGB}{131,207,168}
\definecolor{g9}{RGB}{115,199,157}
\definecolor{g10}{RGB}{98,191,145}
\definecolor{g11}{RGB}{82,185,134}
\definecolor{g12}{RGB}{66,179,123}
\newcommand{\yhy}[1]{\textcolor{red}{#1 [--haoyuan]}}
\renewcommand{\yhy}[1]{#1}

\begin{document}

\maketitle

\begin{abstract}
    Automatic Speech Recognition (ASR) has achieved remarkable success with deep learning, driving advancements in conversational artificial intelligence, media transcription, and assistive technologies. However, ASR systems still struggle in complex environments such as TV series, where multiple speakers, overlapping speech, domain-specific terminology, and long-range contextual dependencies pose significant challenges to transcription accuracy. Existing approaches fail to explicitly leverage the rich temporal and contextual information available in the video. To address this limitation, we propose a Video-Guided Post-ASR Correction (VPC) framework that uses a Video-Large Multimodal Model (VLMM) to capture video context and refine ASR outputs. Evaluations on a TV-series benchmark show that our method consistently improves transcription accuracy in complex multimedia environments.
\end{abstract}

\section{Introduction}
\label{sec:intro}
Automatic Speech Recognition (ASR) has achieved remarkable success with the advent of deep learning, driving advancements in applications such as conversational AI, media transcription, and assistive technologies \cite{DBLP:conf/icassp/GravesMH13}.
% Automatic Speech Recognition (ASR) has achieved remarkable success with the advent of deep learning, advancing applications such as conversational AI, media transcription, and assistive technologies \cite{DBLP:conf/icassp/GravesMH13, DBLP:conf/icml/AmodeiABCCCCCCD16, DBLP:conf/nips/BaevskiZMA20}. ??Deep Speech DBLP:conf/icml/AmodeiABCCCCCCD16
Despite their success, ASR systems still struggle in complex real-world scenarios, particularly in noisy environments \cite{DBLP:journals/ijst/DuaAD23}, domain-specific conversations \cite{DBLP:conf/um/CaoGCSPRKMPD23}, and multimedia content such as TV series or movies \cite{DBLP:conf/jauti/BecaASR18}. These scenarios often involve overlapping speech, domain-specific terminology, and contextual dependencies, making it difficult for ASR systems to rely solely on audio signals to achieve optimal performance.

To address the above challenges, leveraging multimodal contextual information has emerged as a promising direction \cite{DBLP:conf/icassp/NiHC00L024}. 
However, the existing works focusing on utilizing image information to correct the ASR output \cite{DBLP:conf/interspeech/GabeurSN0AS22} \yhy{lack full exploitation of video information \cite{DBLP:conf/cvpr/SeoNS23}}. \yhy{ASR on TV series \cite{DBLP:conf/interspeech/GuinaudeauGS10, DBLP:conf/icassp/KorbarHZ24} is particularly important for accessibility and engagement. In particular, video serves as a powerful complementary modality, providing rich contextual cues that speech alone cannot convey. For example, the video-based context, such as scene elements, character movements, and long-range dependencies inherent in TV series and movies,  can significantly aid in maintaining coherence and provide more useful information for context-dependent disambiguation.}

% DBLP:conf/iberspeech/Perero-Codosero18,

To fully utilize the rich context information from the TV series video, we propose a novel \textbf{training-free multimodal post-ASR correction (VPC) framework} that refines ASR outputs by leveraging contextual content extracted from videos. 
Our framework consists of two stages: 1) ASR Generation, where an ASR model will transcribe the audio; 2) Video-guided Post-ASR Correction. The latter stage is for correcting errors in the generated transcript based on the video. \yhy{Specifically, it consists of two main components: I. The Video-based Contextual Information Extraction module, utilizing the advanced Video-Large Multimodal Model (VLMM) to extract key context information with our devised prompts; II. The Context-aware ASR Correction module aims to correct errors in the generated transcript using the Large Language Model (LLM) and extracted visual context.}

\yhy{We evaluate our approach on a multimodal TV series dataset Violin \cite{DBLP:conf/cvpr/LiuCCGYYL20} with state-of-the-art self-supervised speech models (wav2vec 2.0 \cite{DBLP:conf/nips/BaevskiZMA20}, HuBERT \cite{DBLP:journals/taslp/HsuBTLSM21}, WavLM \cite{DBLP:journals/jstsp/ChenWCWLCLKYXWZ22}), and Conformer \cite{DBLP:conf/interspeech/GulatiQCPZYHWZW20}.} The experiments demonstrate that our method can significantly reduce the Word Error Rate (WER) \yhy{by providing necessary visual context}, yielding a relative 20.75\% improvement on the Violin dataset \cite{DBLP:conf/cvpr/LiuCCGYYL20} for WavLM.

Our contribution can be summarized as follows,
1) To the best of our knowledge, we are the first ones to propose to post-correct for errors in ASR with additional information on video modality.
2) We propose a novel video-guided post-ASR correction method, which can leverage advanced VLMM to extract visual information and LLM to correct the errors in ASR outputs.
And 3) We conduct extensive experiments on the Violin dataset and ASR models. The experimental results show \yhy{that our proposed method provides meaningful visual context that can help improve ASR accuracy under complex multimodal scenarios}.

\begin{figure*}[ht]
    \centering
    \includegraphics[width=1.0\linewidth]{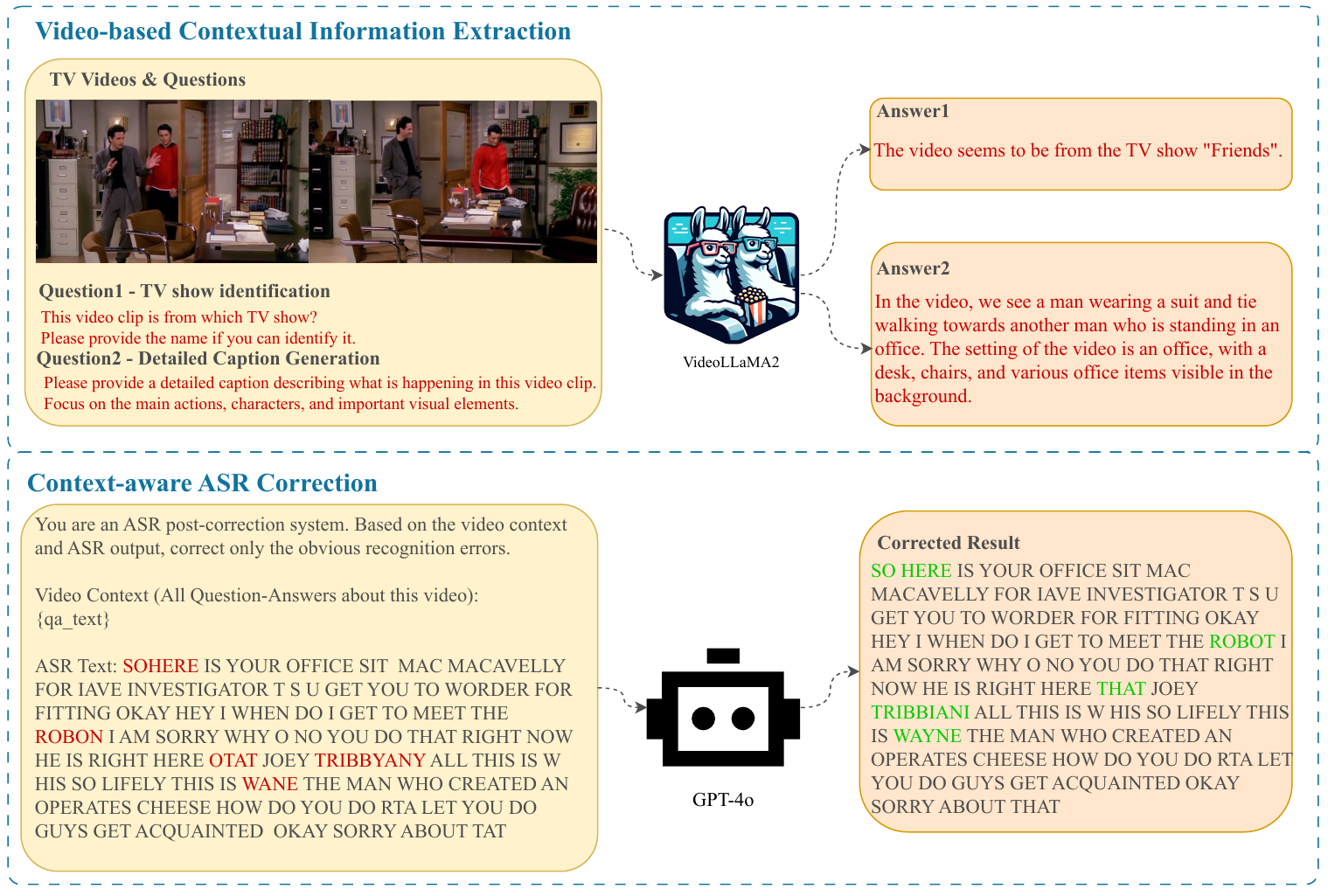}
    \caption{Overview of the proposed Video-Guided Post-ASR Correction (VPC) method. The framework integrates video-based contextual information to refine ASR outputs.}
    \label{fig:vpc_method}
\end{figure*}

\section{Related Work}
{Recent post-ASR correction \cite{ DBLP:journals/corr/abs-2310-13013} aims at correcting the generative errors of a pretrained ASR model using LLMs, with training-free strategy \cite{DBLP:journals/corr/abs-2505-24347} or light additional training \cite{DBLP:conf/emnlp/RadhakrishnanYK23}. 
% By leveraging LLMs, these methods can detect and repair typical errors that occur in complex conversational or noisy settings.
} 
However, real-world ASR tasks often occur in multimodal contexts, where additional information from video can enhance error correction. 
% \yhy{Early research of combining audio-visual modalities focused on activity detection \cite{DBLP:conf/interspeech/BorilSHH10,DBLP:conf/icassp/HasanBSH12, DBLP:journals/ejasp/HasanBSH13}, and later expanded to audio-visual representation learning \cite{DBLP:conf/iclr/ShiHLM22_avhubert} and Audio-Visual ASR \cite{DBLP:conf/interspeech/GabeurSN0AS22, DBLP:conf/cvpr/SeoNS23}.

\yhy{While traditional Audio-Visual Speech Recognition (AVSR) methods, such as AV-HuBERT, have demonstrated success in improving ASR robustness, they primarily rely on low-level sensory fusion \cite{DBLP:conf/interspeech/GabeurSN0AS22, DBLP:conf/cvpr/SeoNS23, DBLP:conf/iclr/ShiHLM22_avhubert}. These models focus on extracting lip-reading cues and facial movements, which require high-resolution, perfectly aligned face tracks to function effectively.}

However, in the complex environment of TV series, such fine-grained visual data is often unavailable due to off-screen speakers, wide shots, or low lighting; at the same time the higher-level context information are overlooked.

% In contrast, our VPC framework focuses on high-level semantic context—such as scene elements, character interactions, and narrative dependencies. By utilizing a Video-Large Multimodal Model (VLMM) to capture this 'cognitive' information, our method can resolve ambiguities (e.g., homophones like 'be hi hat' vs. 'beehive') that are independent of the speaker’s lip movements.

% Audio-visual models \cite{DBLP:conf/interspeech/GabeurSN0AS22, DBLP:conf/cvpr/SeoNS23, DBLP:conf/iclr/ShiHLM22_avhubert} improve recognition by jointly encoding visual and audio features. 

In contrast, the role of video context in post-ASR correction remains underexplored, despite its potential to mitigate errors that persist after decoding.
While these models have demonstrated strong performance, an important research gap remains: the potential role of visual information in further improving ASR performance.
VLMMs \cite{DBLP:journals/corr/abs-2406-07476, DBLP:conf/acl/0001RKK24, DBLP:conf/emnlp/LinYZCNJ024} have excelled in tasks such as video understanding, captioning, and video-based question answering, but their application to ASR correction remains unexplored.
Therefore, different from the existing work, we  propose a novel training-free multimodal post-ASR correct framework, which
integrates visual context from video data to improve the quality of ASR outputs.

% While VLMMs \cite{DBLP:journals/corr/abs-2406-07476, DBLP:conf/acl/0001RKK24, DBLP:conf/emnlp/LinYZCNJ024} have excelled in tasks such as video understanding, captioning, and video-based question answering, their application to ASR correction remains unexplored. To address this gap, we propose a Video-Guided Post-correction framework that leverages video-based contextual information, extending traditional text-based ASR correction systems into the multimodal domain.

\section{Task Formulation}
\label{sec:problem_formulation}
We first introduce the related dataset, then formulate the ASR on TV Series task, and finally introduce the post-ASR correction task.
% \textbf{Dataset.} Suppose we have a training dataset $\mathcal{D}$ composed of $N$ samples, i.e., $\mathcal{D} = \{d_1, d_2, \cdots, d_N\}$. Each sample $d_i = \{A_i, V_i, Y_i\}$, where $A_i = \{a_1^i, a_2^i, \cdots, a_{N_{a_i}}^i\}$ denotes the input audio segmented into $N_{a_i}$ frames, $V_i$ represents the TV series video input, and $Y_i = \{y_1^i, y_2^i, \cdots, y_{N_{y_i}}^i\}$ denotes the target transcription text consisting of $N_{y_i}$ tokens.
\textbf{1) Dataset.} Suppose we have a training dataset $\mathcal{D}$ composed of $N$ samples, i.e., $\mathcal{D} = \{d_1, d_2, \cdots, d_N\}$. Each sample $d_i = \{A_i, V_i, Y_i\}$, where $A_i$ denotes the input audio segmented into $N_{a_i}$ frames, $V_i$ represents the TV series video input, and $Y_i$ denotes the target transcription text consisting of $N_{y_i}$ tokens. \textbf{2) ASR on TV Series.} Based on these training samples, the target of ASR is to learn an ASR model $\mathcal{F}$ that maps the given audio input to its corresponding transcription as follows:
\begin{equation}
    \hat{Y}_i = \mathcal{F}(A_i\mid \Theta),
\end{equation}
where $\Theta$ represents the set of parameters to be learned for the ASR model $\mathcal{F}$, and $\hat{Y}_i$ is the predicted transcription of the $i$-th sample.
\textbf{3) Video-guided post-ASR correction.} Based on the generated transcription from the ASR model, the video-based correction task aims to identify and repair errors in the generated transcription as follows,
\begin{equation}
    \bar{Y}_i = \mathcal{M}(\hat{Y}_i, V_i),
\end{equation}
where $\bar{Y}_i$ is the corrected transcription based on the $\hat{Y}_i$ and video $V_i$. $\mathcal{M}$ is our devised post-ASR correction method.

\section{Methodology}
\label{sec:method}
We propose a novel multimodal post-ASR correction framework that integrates visual context from video data to improve the quality of ASR outputs. The framework consists of two main stages: \textbf{ASR Generation Stage} and \textbf{Video-guided Post-ASR Correction Stage}.

\subsection{ASR Generation}
Despite the compelling success recent ASR systems have achieved in several datasets \cite{DBLP:conf/icassp/KahnRZKXMKLCFLS20, DBLP:journals/corr/abs-1207-0580, DBLP:conf/icassp/PanayotovCPK15}, they still struggle with context-dependent disambiguation, phonetic approximation errors, and homophone errors. For example, when transcribing TV series dialogues, it may misrecognize \textit{Joey Tribbiani} as \textit{Joey Tribbyany}. This error occurs because the model lacks sufficient world knowledge to correctly infer the proper spelling of ambiguous words that is context-dependent, especially those that are uncommon or do not frequently appear in the training data. Therefore, we devised a video-guided post-ASR correction method to correct the output for ASR models. In this stage, we feed the ASR model with audio and get the corresponding generated transcript $\hat{Y}$.

% However, wav2vec 2.0 does not include a built-in language model, which can result in semantic errors.

% \subsubsection{Whisper}
% Whisper is a Transformer-based encoder-decoder model trained on a large-scale multilingual speech dataset. It converts audio into log-Mel spectrograms and uses supervised learning to perform ASR. Whisper integrates both acoustic and language modeling, making it suitable for tasks involving complex semantics and multilingual settings.

\subsection{Video-guided Post-ASR Correction}
To enhance ASR output quality, we propose a Video-guided Post-ASR Correction (VPC) method that leverages contextual information extracted from video content. The framework is illustrated in Figure~\ref{fig:vpc_method}, and consists of two main components: Video-based Contextual Information Extraction and Context-aware ASR Correction.

\subsubsection{Video-based Contextual Information Extraction}
% \yhy{ORIGINAL}Considering that the videos usually contain rich contextual information (e.g., scenes, actions, and characters in movies),  we propose to extract visual information as additional information to correct the errors in the generated transcript. 
% Since the VLMM \cite{DBLP:journals/corr/abs-2406-07476, DBLP:conf/acl/0001RKK24, DBLP:conf/emnlp/LinYZCNJ024} have demonstrated impressive performance in tasks such as video understanding, caption generation, and video-based question answering by integrating visual and textual data,
% % shows excellent performance in video understanding tasks, such as Video Question Answering \cite{}, 
% we resort to the VLMM to capture the rich semantic context from the video.\yhy{ORIGINAL}

{Given VLMMs' impressive performance in video understanding, we resort to VLMM to capture the rich semantic context (e.g., scenes, actions, and characters) in video.}
Specifically, we extract the fine-grained information from the video with a Question-Answering (QA) format. We devised two questions to fully mine the information from the video: 1) TV Show Recognition and 2) Fine-grained Video Description. 
The potential reason we devised these two questions is: 1) the TV show name can be utilized to retrieve knowledge related to the TV show, such as character names, and the QA-based information extraction method could mine this kind of information from videos; 2) ASR models sometimes generate errors that are unrelated to the video content or topic, and fine-grained video description could be used to correct such errors.

% given a video $V$, the model extracts scene-relevant details that include entities, actions, and environmental descriptions.

We devise two prompts for the two kinds of questions, which can be used to prompt VLMM for rich context information. In particular, we feed the video $V$ and prompts to VLMM and get the corresponding response as follows,
\begin{align}
    C_1 = \operatorname{VLMM}(V, P_1),\\
    C_2 = \operatorname{VLMM}(V, P_2),
\end{align}
where $P_1$ and $P_2$ are the prompts for two kinds of questions, as shown in Figure \ref{fig:vpc_method}. 
$C_1$ and $C_2$ are rich context information generated by $\operatorname{VLMM}$ according to the input video and prompts.

% descriptions $C$, represented as textual tokens. 
% This textual output provides meaningful semantic context for the ASR correction stage, ensuring alignment between transcriptions and video content. Formally, for each video $V$, we extract:
% \begin{equation}
% C = \operatorname{VLMM}(V, P),
% \end{equation}
% where $C = \{c_1, c_2, \ldots, c_{N_c}\}$ represents the sequence of extracted contextual tokens, and $N_c$ is the total number of tokens. $P$ is the prompt which is shown in Figure \ref{fig:vpc_method}.

\subsubsection{Context-aware ASR Correction} 
In this stage, we aim to correct the ASR outputs using the extracted visual context information $C_1$ and $C_2$. A straightforward approach would be to train a dedicated model for the correction task; however, this requires extensive data collection and model training, which are both time-consuming and resource-intensive. Recently, Large Language Models (LLMs) have shown excellent performance in language understanding and generation; we resort to LLMs to correct errors in transcripts with context information.

% To refine the ASR outputs, we integrate the extracted visual context information $C$ with the initial ASR predictions. The goal is to correct common errors such as insertions, deletions, and substitutions, ensuring the final transcription aligns with the video’s semantic meaning.

Specifically, we feed the LLM with a task instruction, the extracted context information $C_1$ and $C_2$ from the input video as follows,
\begin{equation}
\bar{Y} = \operatorname{LLM}(\hat{Y}, C_1, C_2, T),
\end{equation}
where $T$ is the task instruction, as shown in Figure \ref{fig:vpc_method}. $\hat{Y}$ is the initial transcription generated by an ASR model (e.g., wav2vec 2.0). $\bar{Y}$ is the corrected transcript by the $\operatorname{LLM}$.

% To ensure the quality of the correction, we train $\mathcal{G}$ using a supervised loss that minimizes the word error rate (WER) between the corrected output $\hat{Y}$ and the ground truth transcription.

\begin{table*}[!h]
\centering
\resizebox{\textwidth}{!}{
\begin{tabular}{cc|cc|ccc}
\toprule
\multirow{2}{*}{Model} & \multirow{2}{*}{Raw ASR} &
\multicolumn{5}{c}{Post-ASR Correction} \\
\cmidrule(lr){3-7}
 & & GPT (w/o visual context) & Raw-Impro↑ & \textbf{VPC} & Raw-Impro↑ & GPT-Impro↑ \\
\midrule
 wav2vec2-large  & 29.17 & 29.28 & -0.38\% & \textbf{25.36} & \cellcolor{g7}{13.06\%} & \cellcolor{g7}{13.39\%} \\
 HuBERT-large    & 26.40 & \textbf{25.73 }&  \cellcolor{g2}{2.54\%} & \textbf{23.27} & \cellcolor{g6}{11.86\%} &  \cellcolor{g5}{9.56}\% \\
 WavLM-large     & 29.83 & 29.45 &  \cellcolor{g1}{1.27\%} & \textbf{23.64} & \cellcolor{g11}{20.75\%} & \cellcolor{g10}{19.73}\% \\
 Conformer-large & 22.66 & 22.14 &  \cellcolor{g2}{2.29}\% & \textbf{20.97} &  \cellcolor{g4}{7.46}\% &  \cellcolor{g3}{5.28\%} \\
\bottomrule
\end{tabular}}
\caption{WER comparison of models on the Violin-TV subset. GPT uses no visual context prompt, while VPC augments GPT with VLMM-based visual context. Raw-Impro and GPT-Impro are the relative WER improvement between our VPC and Raw ASR and GPT (w/o visual context), respectively.}
\label{tab:long_clips_results}
\end{table*}

\section{Experiments}

% \subsection{Experimental Setup}

\subsection{Dataset.} 
%We evaluate our approach on a TV subset of the Violin dataset \cite{DBLP:conf/cvpr/LiuCCGYYL20}, which provides multimodal content including audio, video, and subtitles from TV series. We only choose video clips from TV series because they contain a higher proportion of speech compared to other types of content. Some clips from Violin consist solely of video with background music, while others feature English subtitles but are spoken in different languages. Therefore, we aim to exclude these clips from our selection.

We evaluate our approach on the Violin dataset \cite{DBLP:conf/cvpr/LiuCCGYYL20}, which provides multimodal content including audio, video, and subtitles from TV series, movies, and YouTube clips. While the dataset provides diverse clips, some contain only background music without speech, while others include English subtitles but are spoken in different languages. 
% To ensure our focus on speech-rich content, we exclude these noisy clips.
% \lj{how did you remove them?}

To ensure the focus on ASR, we create a Violin-TV subset by selecting clips from only TV series in which English is the primary spoken language. The subset comprises 10,003 TV clips, including the corresponding audio, transcript, and video.
The average duration of the subset is  32.4 seconds, with an average speech density of 74.9\%. With a total duration of 90.027 hours, we allocate a 72-hour training set with 7,983 clips, a 9-hour validation set with 1,007 clips and a 9-hour testing set with 1,013 clips.
% audio from TV clips with an average duration of 32.4 seconds, paired with their original video frames. 
We use the original transcript in the dataset as the ground-truth transcript.

% We evaluate our approach on a subset of Violin \cite{DBLP:conf/cvpr/LiuCCGYYL20}, which provides multimodal content including audio, video, and subtitles from TV series, movies, and YouTube clips. Violin-TV contains only clips from four TV series in which English is the primary language. The Violin-TV subset comprises 10,003 TV clips, including the corresponding audio, \yhy{subtitle}, and video.
% The average duration of the subset is  32.4 seconds, with an average speech density of 74.9\%. We cleaned the original transcript in the dataset and used it as the ground-truth transcript.

% With a total duration of 90.027 hours, we allocate a 72-hour training set with 7,983 clips, a 9-hour validation set with 1,007 clips and a 9-hour testing set with 1,013 clips.

\subsection{Implementation Details.}

To assess our VPC's effectiveness in correcting transcription, we selected wav2vec 2.0, HuBERT and WavLM, and Conformer as our ASR models.  All models are pretrained on Librispeech-960h \cite{DBLP:conf/icassp/PanayotovCPK15}. We finetuned wav2vec 2.0, HuBERT and WavLM with CTC loss, and finetuned Conformer with RNN-T loss on the 72-hour training set.
% We apply our VPC method to the generated transcripts from state-of-the-art models s. 
In our VPC method, we apply VideoLLaMA2 \cite{DBLP:journals/corr/abs-2406-07476} as our VLMM, and GPT-4o as our LLM. For the evaluation metric, we use WER. 

We considered including state-of-the-art joint Audio-Visual ASR (AV-ASR) models, such as AV-HuBERT \cite{DBLP:conf/iclr/ShiHLM22_avhubert}, as baselines. However, our preliminary experiments of AV-HuBERT on the Violin dataset showed that these models performed poorly, yielding a 78.3\% WER. This failure is primarily due to the inconsistent face resolution and frequent off-screen speakers inherent in TV series, which prevent these models from consistently finding or resolving the lip-movement cues they rely on. 

We didn't include AV-HuBERT as a baseline because it relies on high-resolution, aligned face tracks to function, meaning that it is ill-suited for this complex domain. Consequently, we focus our comparison on post-ASR correction methods that leverage high-level semantic context, which is more robust to these visual challenges.

% \yhy{We use 3 SSL speech models, wav2vec 2.0, HuBERT and WavLM, finetuned with CTC loss, and a Conformer finetuned with RNN-T loss as our ASR models. All models are pretrained on Librispeech-960h.} 

% \yhy{To further justify the effectiveness of the Video-based Contextual Information Extraction module, we add another baseline which prompts GPT-4o to correct the generated ASR result from the ASR model without visual context.}

% \textit{``You are an ASR post-correction system. ASR outputs are from famous TV series. Please try to recognize the TV series name and correct only the obvious recognition errors."}

% To verify the effectiveness of our method in extensive sceneries, we further fine-tune baselines with two different settings in terms of data size \yhy{TODO: Update Dataset Usage -- we don't use the 1 hour fine-tuning subset now}:
% 1) FT 1-hour: The model is fine-tuned on 1-hour subset randomly sampled from the training data of Violin-TV. 
% 2) FT 72-hour: The model is fine-tuned on all the training data in Violin-TV. We further apply our method to model variants of the two settings.

\subsection{Results Analysis}
We present the WER results of different models on the Violin-TV subset in Table~\ref{tab:long_clips_results}. 
From this Table, we have several observations:
\yhy{
1) \textbf{Proposed VPC constantly improves all models.} With additional visual context, the proposed VPC achieves 20.75\% relative WER reduction on WavLM-Large, 13.06\% on wav2vec 2.0, 11.86\% on HuBERT and 7.64\% on Conformer-Large with RNN-T loss. This highlights that leveraging video information using QA and helping refine ASR output is effective and robust.
2) \textbf{VPC is Training-free and generalizable.} Our proposed VPC needs no further training with ASR models. The improvements across different models suggest that it is easy to integrate and generalize.
3) \textbf{GPT-4o alone without visual information performs worse than our VPC method}. Using GPT-4o directly yields negligible or even negative effects (e.g., –0.38\% relative on wav2vec2-large) compared to the original ASR output. This indicates that text-only LLMs cannot reliably handle ASR correction in complex multimodal settings, where cross-modal dependencies and long-range context play an essential role.
}

\begin{figure}[t]
    \centering
    \includegraphics[width=1.0\linewidth]{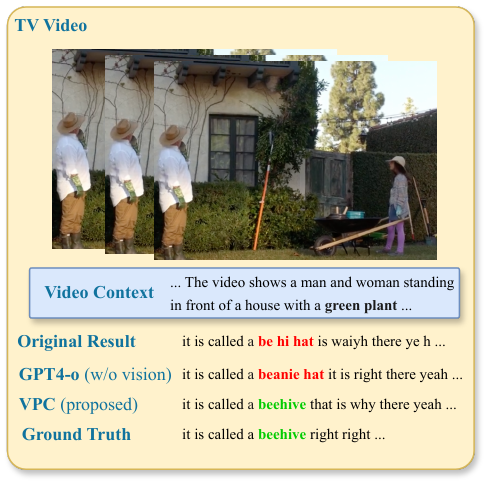}
    \caption{Comparison between results generated by WavLM and its corresponding variant augmented with VPC.}
    \label{fig:case}
\end{figure}

% Unlike AV-ASR systems that needs jointly training the AV encoder, our proposed VPC model needs no training with a fine-tuned ASR. The improvements shown across different models suggest that it is easy to intergrate and generalize.
\subsection{Prompt Sensitivity Analysis}
% 4.X. Prompt Sensitivity Analysis
A common concern with Large Multimodal Models (LMMs) and LLMs is their sensitivity to prompt engineering, which can lead to ``hallucinations" or inconsistent performance. 
To evaluate the robustness of our VPC framework, we conducted a pilot study on a randomly selected 100-clip subset of the Violin dataset to evaluate three prompt strategies, we conducted a sensitivity analysis by testing three different prompt templates for the VLMM: (1) \textit{Coarse-QA}, asking for the name of the show and provide background info, (2) \textit{Fine-QA}, asking for more detailed context of the shown clip, include scenes, actor (3) \textit{All-QA} (chosen) that asks both Coarse and Fine QA.

\begin{table}[t]
  \caption{Prompt sensitivity analysis (WER \%) on a 100-clip pilot subset.}
  \label{tab:sensitivity}
  \centering
  \resizebox{\columnwidth}{!}{%
  \begin{tabular}{lccc}
    \toprule
    \textbf{Prompt Strategy} & \textbf{Wav2Vec 2.0} & \textbf{HuBERT} & \textbf{WavLM} \\
    \midrule
    Coarse-QA               & 25.32              & \textbf{22.90}             & 23.37           \\
    \textit{Fine-QA}           & 25.41              & 23.15             & 23.72           \\
    \rowcolor{g1}
    \textbf{\textit{All-QA} (Chosen)} & \textbf{25.30}     & 22.93    & \textbf{22.54}  \\
    \bottomrule
  \end{tabular}
  }
\end{table}

As shown in Table~\ref{tab:sensitivity}, we observed marginal changes in WER across different prompt strategies, indicating the framework's robustness to specific phrasing. Interestingly, without the broader scene context, the performance of \textit{Fine-QA} (detailed descriptions) was slightly inferior to \textit{Coarse-QA} (overall context). Consequently, we selected the \textit{All-QA} strategy in the main results for all subsequent experiments, as it effectively captures both high-level situational context and fine-grained visual details, providing the most stable performance gains across backbones.

\subsection{Case Study}
We further show a case study about the comparison between results generated by WavLM and its corresponding variant augmented with
our method. In this sample, our proposed model revises the error words ``a be hi hat'' into the correct word ``a beehive''. This showcases ability of the context-dependent disambiguation since the video content provides additional information to facilitate the LLM to revise the potential error by reasoning. This further verifies the effectiveness of our method.

\section{Conclusion}
% \yhy{Needs revision after results analysis}
\yhy{In this work, we introduced a novel video-guided post-ASR correction (VPC) method to enhance ASR performance in complex multimedia environments. By effectively leveraging  Video-Large Multimodal Model (VLMM) to extract rich contextual information from videos and integrating it with Large Language Model (LLM), our method significantly reduces ASR errors on a multimodal TV benchmark. Experimental results demonstrate the effectiveness and generalizability of our method. Moving forward, our work paves the way for more robust multimodal ASR correction strategies that integrate deeper video understanding, making ASR systems more adaptable and reliable in real-world multimedia applications.}

% our method significantly reduces ASR errors on a multimodal TV benchmark. These results highlight the promise of lightweight multimodal correction strategies for making ASR more robust in real-world multimedia environments.

% {\color{blue}
% \section{Generative AI Use Disclosure}
% The extent of Generative AI use must be disclosed. This section may be in the 5th or 6th pages of regular papers, or the 9th or 10th pages of long papers.  ISCA policy says: \textit{All (co-)authors must be responsible and accountable for the work and content of the paper, and they must consent to its submission. Any generative AI tools cannot be a co-author of the paper. They can be used for editing and polishing manuscripts, but should not be used for producing a significant part of the manuscript}.}
\section{Generative AI Use Disclosure}
During the preparation of this work, all (co-)authors only used Gen AI tools to review and make corrections on grammar and choice of words. After the editing and polishing, the (co-)authors reviewed and edited the content as needed and take full responsibility for the content of the paper.

% \bibliographystyle{IEEEtran}
% \bibliography{mybib}

\bibliographystyle{IEEEtran}
\bibliography{bib/refs,bib/audio_video,bib/post_correction}
% \bibliography{bib/refs, bib/audio_video, bib/post_correction}

\end{document}